\documentclass{article}
\usepackage{epsfig}

\date{\today}

\usepackage[hang,nooneline]{subfigure}
\usepackage[intlimits,tbtags]{amsmath} 
\usepackage{amsfonts,amssymb}
\usepackage{amsthm}
\usepackage{amscd}

\begin{document}
\title{Holographic superfluid/fluid/insulator phase transitions in 2+1 dimensions}
\author{{\large Yves Brihaye \footnote{email: yves.brihaye@umons.ac.be} }$^{\ddagger}$ and 
{\large Betti Hartmann \footnote{email: b.hartmann@jacobs-university.de}}$^{\dagger}$
\\ \\
$^{\ddagger}${\small Physique-Math\'ematique, Universite de
Mons-Hainaut, 7000 Mons, Belgium}\\ 
$^{\dagger}${\small School of Engineering and Science, Jacobs University Bremen, 28759 Bremen, Germany}  }

\date{}
\newcommand{\dd}{\mbox{d}}
\newcommand{\tr}{\mbox{tr}}
\newcommand{\la}{\lambda}
\newcommand{\ka}{\kappa}
\newcommand{\f}{\phi}
\newcommand{\vf}{\varphi}
\newcommand{\F}{\Phi}
\newcommand{\al}{\alpha}
\newcommand{\ga}{\gamma}
\newcommand{\de}{\delta}
\newcommand{\si}{\sigma}
\newcommand{\bomega}{\mbox{\boldmath $\omega$}}
\newcommand{\bsi}{\mbox{\boldmath $\sigma$}}
\newcommand{\bchi}{\mbox{\boldmath $\chi$}}
\newcommand{\bal}{\mbox{\boldmath $\alpha$}}
\newcommand{\bpsi}{\mbox{\boldmath $\psi$}}
\newcommand{\brho}{\mbox{\boldmath $\varrho$}}
\newcommand{\beps}{\mbox{\boldmath $\varepsilon$}}
\newcommand{\bxi}{\mbox{\boldmath $\xi$}}
\newcommand{\bbeta}{\mbox{\boldmath $\beta$}}
\newcommand{\ee}{\end{equation}}
\newcommand{\eea}{\end{eqnarray}}
\newcommand{\be}{\begin{equation}}
\newcommand{\bea}{\begin{eqnarray}}

\newcommand{\ii}{\mbox{i}}
\newcommand{\e}{\mbox{e}}
\newcommand{\pa}{\partial}
\newcommand{\Om}{\Omega}
\newcommand{\vep}{\varepsilon}
\newcommand{\bfph}{{\bf \phi}}
\newcommand{\lm}{\lambda}
\def\theequation{\arabic{equation}}
\renewcommand{\thefootnote}{\fnsymbol{footnote}}
\newcommand{\re}[1]{(\ref{#1})}
\newcommand{\R}{{\rm I \hspace{-0.52ex} R}}
\newcommand{\N}{{\sf N\hspace*{-1.0ex}\rule{0.15ex}%
{1.3ex}\hspace*{1.0ex}}}
\newcommand{\Q}{{\sf Q\hspace*{-1.1ex}\rule{0.15ex}%
{1.5ex}\hspace*{1.1ex}}}
\newcommand{\C}{{\sf C\hspace*{-0.9ex}\rule{0.15ex}%
{1.3ex}\hspace*{0.9ex}}}
\newcommand{\eins}{1\hspace{-0.56ex}{\rm I}}
\renewcommand{\thefootnote}{\arabic{footnote}}

\maketitle

\bigskip

\begin{abstract}
We study the breaking of an Abelian symmetry close to the horizon of
a black string as well as close to the tip of a 
solitonic, cigar-shaped solution in (3+1)-dimensional Anti--de Sitter space-time. We use these
solutions to describe holographic superfluids away from the probe limit, i.e. taking backreaction into account.
We observe that
up to four phases exist in this model representing the duals of black string solutions with and without scalar hair and
solitonic, cigar-shaped solutions with and without scalar hair, respectively. We construct the full phase diagram that describes the phase
transitions between fluids and superfluids, between insulators and superfluids as well as between insulators and fluids.
In the probe limit the phase transition from fluids to black string superfluids changes from being second order to first order
for sufficiently large values of the superfluid velocity and/or the angular momentum of the dual black string.
We find that if we take backreaction into account phase transitions that are first order
for weak backreaction become again second order for sufficiently strong backreaction.
Moreover, we find a new type of insulator/superfluid phase transition for strong backreaction and vanishing superfluid velocity
as well as a new type of fluid/superfluid phase transition that exists only for non-vanishing superfluid velocity.
\end{abstract}
\medskip
\medskip

\section{Introduction}
The gravity--gauge theory duality \cite{ggdual} has attracted a lot of attention in the past years. The most famous
example is the AdS/CFT correspondence \cite{adscft} which states that a gravity theory in a $d$-dimensional
Anti-de Sitter (AdS) space--time is equivalent to a Conformal Field Theory (CFT) on the $(d-1)$-dimensional boundary of AdS.

Recently, this theory has been used to describe so-called holographic superconductors and superfluids with the help of black holes in
higher dimensional space--time \cite{gubser,hhh,reviews}.  The main idea is that 
close to the horizon of a black hole in AdS space--time the Abelian symmetry can be spontaneously broken.
The reason for this is that
close to the horizon of the black hole the effective mass of the scalar field can become
negative with masses below the Breitenlohner--Freedman bound \cite{bf} such that the scalar
field becomes unstable and possesses a non--vanishing value on and close to the horizon
of the black hole.
The local bulk U(1) symmetry  is associated to a global U(1) symmetry on the AdS boundary and
the value of the scalar field on the AdS boundary  with the corresponding condensate in the dual theory.
The Hawking temperature of the black hole is identified with the temperature of the dual theory.
For temperatures above a critical value $T_c$ the black hole possesses no scalar hair and is the gravity dual of a conductor, while for 
temperatures below $T_c$ 
the black hole has scalar hair and corresponds to a superconductor. These models can be extended to describe holographic fluid/superfluid
phase transitions by making the spatial component of the U(1) gauge field in the bulk theory non-vanishing which
corresponds to non-vanishing superfluid currents on the AdS boundary \cite{hks,superfluids,hartmann_brihaye}.
It was
shown that for sufficiently large superfluid velocity the phase transition changes from being 2nd to 1st order. 
Interestingly, this changes when taking strong backreaction of the space--time into account. In \cite{sonnerwithers} it was shown that in this case the phase transition remains
2nd order for all values of the superfluid velocity.

Interestingly, the model used to describe conductor/superconductor and fluid/superfluid phase transitions, respectively
by using black holes in higher-dimensional AdS can be modified to describe insulator/superconductor phase
transitions by considering soliton solutions in higher-dimensional AdS. This was done in \cite{nrt} and \cite{horowitz_way}
in (4+1)-dimensional AdS in the probe limit and taking backreaction into account, respectively.
The AdS soliton is related to the black hole by a double Wick rotation. Moreover, one of the coordinates is compactified to a circle.
The AdS soliton has been used before in the context of the description of a confining vacuum 
in the dual gauge theory \cite{witten2,horowitz_myers} since it possesses a mass gap. While for spherically symmetric
black holes in AdS there is the Hawking-Page phase transition from the AdS black hole to global AdS space-time when
lowering the temperature \cite{hawking_page}, this is different for AdS black holes with Ricci-flat horizons used in the description of holographic superconductors.
In that case, there is a phase transition between the AdS black hole and the AdS soliton \cite{ssw} which was interpreted as
a confining/deconfining phase transition in the dual gauge theory. Now taking the viewpoint of  condensed
matter this phase transition describes a 1st order insulator/conductor phase transition. 
To complete the picture it was realized in \cite{nrt} that one can add a chemical potential to the AdS soliton.
Increasing the chemical potential $\mu$ the AdS soliton becomes unstable to the formation of scalar hair above some
critical value $\mu_{cr}$. This was interpreted as a 2nd order insulator/superconductor phase transition that is possible
even at zero temperature. The study done in the probe limit in \cite{nrt} was extended to take backreaction into account \cite{horowitz_way}. 
Phase diagrams were constructed and it was shown that there is a new type of
phase transition from a superconductor to an insulator for sufficiently strong backreaction. 
  
In this paper we want to extend these results to (3+1)-dimensional AdS. Moreover, we add a non-vanishing
spatial component of the U(1) gauge field and consider gravity solutions that possess a non-vanishing angular
momentum. Our aim is to describe insulator/fluid/superfluid phase transitions by studying the breaking of an Abelian symmetry close to the 
horizon of a black string as well as close to the tip of a solitonic, cigar-shaped
solution in (3+1)--dimensional AdS space--time. The two solutions are related to each other by a double Wick rotation.
The static, i.e. non-rotating black string solutions correspond to the planar black hole solutions frequently used in the
discussion of holographic superconductors and superfluids in 2+1 dimensions. The only difference is
that in our case one of the dimensions is compactified, i.e. the boundary theory lives on $\mathbb{R}^2\times S^1$.
The fact that one of the dimensions is compact is crucial in our case since it will lead
to the existence of up to four phases. This has already been observed in \cite{nrt,horowitz_way}, where static black holes and solitons in (4+1)-dimensional
AdS were discussed. Uncharged, rotating black string solutions have first been discussed in \cite{lemos1,lemos2}, while
charged generalizations were given in \cite{lemos3}. In \cite{dehghani2} a dilatonic rotating and charged black string was considered.
The thermodynamics of these objects has also been studied \cite{dehghani3}. In \cite{hartmann_brihaye} uncharged rotating
black strings were used to describe a holographic fluid/superfluid phase transition in the probe limit. It was found
that for sufficiently large ratio between the superfluid velocity and the chemical potential the phase transition becomes
1st order, while it is 2nd order for small ratio. Though there is no rotation on the AdS boundary, the angular momentum of the space-time
influences the holographic description. It was found that
when increasing the angular momentum the phase transition again becomes 1st order, while it is 2nd order in the static limit.

Our paper is organized as follows: in Section 2, we give the model including the equations
and boundary conditions. In Section 3, we discuss black string and solitonic solutions without scalar hair
that describe holographic fluids and insulators in the dual theory, respectively. In Section 4, we discuss the free energy.
In Section 5, we give our
numerical results for solutions with scalar hair. In particular, we present our results for the formation of scalar
hair on black strings which corresponds to a holographic fluid/superfluid phase transition as well as for the formation
of scalar hair on the tip of the solitonic, cigar-shaped solution which describes a holographic insulator/superfluid
phase transition. We also construct the full phase diagram in Section 5. Section 6 contains our conclusions. 

\section{The Model}
In this paper, we are studying the formation of scalar hair on a rotating black string and a 
rotating solitonic, cigar-shaped solution, respectively, in $(3+1)$--dimensional Anti--de Sitter 
space--time taking backreaction of the space-time into
account. The action reads
\begin{equation}
\label{action}
 S=\int d^4 x \sqrt{-g}\left(R + \frac{6}{\ell^2} + 16\pi G{\cal L}_m\right)  \ ,
\end{equation}
where $\Lambda=-3/\ell^2$ is the cosmological constant and $G$ is Newton's constant.
The matter Lagrangian density ${\cal L}_m$ of a complex scalar field minimally coupled to a U(1) gauge field reads
\begin{equation}
{\cal L}_m=-\frac{1}{4} F_{\mu\nu} F^{\mu\nu} - \left(D_{\mu}\Psi\right)^* D^{\mu}\Psi - m^2 \Psi^*\Psi  \ \ \ \ , \ \  \mu,\nu=0,1,2,3  \ ,
\end{equation}
where $F_{\mu\nu} =\partial_{\mu} A_{\nu} - \partial_{\nu} A_{\mu}$ is the field strength tensor and
$D_{\mu}\Psi=\partial_{\mu} \Psi - ie A_{\mu} \Psi$ is the covariant derivative.
$e$ and $m^2$ denote the gauge coupling and the mass of the scalar field $\Psi$, respectively.
The equations of motion resulting for the variation of this action read
\begin{eqnarray}
 \frac{1}{\sqrt{g}} D_{\mu}\left(\sqrt{-g}g^{\mu\nu} D_{\nu}\Psi\right) &=& m^2 \Psi \ \ , \\
 \frac{1}{\sqrt{g}} \partial_{\mu}\left(\sqrt{-g}g^{\mu\nu} g^{\rho\sigma} F_{\nu\sigma}\right)&=&e g^{\rho\lambda} J_{\lambda} \ , \\
G_{\mu\nu}-\frac{3}{\ell^2} g_{\mu\nu}=8\pi G T_{\mu\nu}  \ ,
\end{eqnarray}
where $J_{\lambda}=i\left[\Psi^* \left(D_{\lambda}\Psi\right) - \Psi \left(D_{\lambda} \Psi\right)^* \right]$ is the
4--current and $T_{\mu\nu}$ is the energy-momentum tensor given by
\begin{equation}
 T_{\mu\nu}= F_{\mu\lambda} F_{\nu}^{\lambda} + D_{\mu} \Psi (D_{\nu} \Psi)^* + D_{\nu} \Psi (D_{\mu} \Psi)^*
+ g_{\mu\nu} {\cal L}_m \ . 
\end{equation}

\subsection{The Ansatz}
Choosing cylindrical coordinates $(r,\chi,z)$ the Ansatz 
for the gauge field $A_{\mu}$ and complex scalar field $\Psi$ reads
\be
\label{ansatz_gauge}
    A_{\mu}dx^{\mu} = \phi(r) dt + A(r) d\chi \ \ \ , \ \ \Psi= e^{-i\omega t + i n \chi} \psi(r)  \ \ , 
\ee
where $n\in {\mathbb{Z}}$, $\omega$ is a constant and $\chi$ is a periodic coordinate. 
Using the U(1) gauge transformation the scalar field can be chosen to be real as long as $e\neq 0$, where $e=\infty$ corresponds
to the probe limit. For $e=0$ the scalar
field $\Psi$ decouples from the electromagnetic field and the phase cannot be gauged away. Following \cite{horowitz_way} the solutions
would then correspond to planar boson stars that in the dual theory would describe Bose-Einstein condensates of glueballs.
In this paper we will always choose $e > 0$. We can perform a rescaling of the type 
$\psi \rightarrow \psi/e$, $\phi\rightarrow \phi/e$, $A\rightarrow A/e$. The equations of motion
then depend only on the mass $m^2$ of the scalar field and on the parameter
$\alpha=8\pi G/e^2$. Hence, the probe limit $e\rightarrow \infty$ corresponds to
$\alpha=0$. Increasing $\alpha$ corresponds to increasing the backreaction of the space-time.

In the following, we would like to discuss two different space-times: (i) a black string space-time that possesses a horizon
and (ii) a solitonic, cigar-shaped space-time that is globally regular. In both cases, we will
discuss static and stationary solutions. In the dual
description, the formation of scalar hair on the black string corresponds to a 
fluid/superfluid phase transition, while the formation of scalar hair on the cigar solution will
describe an insulator/superfluid phase transition. In addition there is a phase transition from
the black string without scalar hair to a soliton solution without scalar hair which in the dual
description corresponds to an insulator/fluid phase transition. Note that the richness of the phase
space in our model results from the fact that one of the coordinate directions is compactified to a circle.
This is very similar to what happens in (4+1)-dimensional AdS space-time \cite{nrt,horowitz_way}.
\subsubsection{Black string solutions}

The Ansatz for the metric of a black string solution reads with $\chi=\varphi$
\begin{equation}
\label{newmetric}
 (ds^2)_{\rm BS} = -b(r)dt^2 + \frac{1}{f(r)} dr^2  + r^2 \left(g(r) dt - d\varphi\right)^2 + p(r) dz^2 \  .
\end{equation} 
This solution has a horizon at $r=r_h$ with $f(r_h)=0$ and $b(r_h)=0$. 
The static limit corresponds to $g(r)\equiv 0$. The Killing vector that is null on the horizon $r_h$ is given
by 
\begin{equation}
 \chi=\partial_t + g_h \partial_{\varphi}
\end{equation}
where $g_h=g(r_h)$ corresponds to the angular velocity $(d\varphi/dt)$ on the horizon.

The temperature of the black hole is given by
$T=\frac{\kappa}{2\pi}$, 
where $\kappa$ is the surface gravity with
\begin{equation}
 \kappa^2=\left. -\frac{1}{2} (D_{\mu} \chi_{\nu}) (D^{\mu} \chi^{\nu})\right\vert_{r_h} \ .
\end{equation}
Using that $f=f'(r_h)(r-r_h)+..$ and $b=b'(r_h)(r-r_h)+..$
close to the horizon $r_h$, where here and in the following the prime denotes the derivative with respect to $r$, we find that
\begin{equation}
\label{t_bs}
 T=\frac{\sqrt{f'(r_h)b'(r_h)}}{4\pi}  \ .
\end{equation}

\subsubsection{Soliton solutions}

The metric for a solitonic, cigar-shaped solution can be obtained from the static limit $g(r)\equiv 0$ of (\ref{newmetric}) 
by a double Wick rotation $t\rightarrow iz$, $z\rightarrow it$ and then by
exchanging the role of the coordinates $\varphi$ and $z$, i.e.
making the $\varphi$ direction non-periodic and calling this coordinate $z$ and making the $z$-direction periodic
and calling the new coordinate $\eta$. Note that $\eta$ here is the periodic coordinate $\chi$ (see (\ref{ansatz_gauge})). 
Hence the metric for a static cigar solution reads
\begin{equation}
\label{cigar_static}
 (ds^2)_{\rm cigar}^{\rm static} = -p(r) dt^2 +\frac{1}{f(r)} dr^2 + r^2 dz^2 + b(r) d\eta^2  \ .
\end{equation}
Then, for a stationary cigar solution that rotates around the $z$-axis the metric reads
\begin{equation}
 (ds^2)_{\rm cigar}^{\rm stationary} = -p(r) dt^2 +\frac{1}{f(r)} dr^2 + r^2 dz^2 + b(r)\left(g(r)dt- d\eta\right)^2  \ .
\end{equation}
The tip of the cigar at $r=r_0$ is given by the requirement $b(r_0)=0$ and $f(r_0)=0$. In order to avoid a conical singularity
at $r=r_0$ we have to require that the $\eta$-coordinate is periodic, i.e. we ``smooth out'' the tip of the cylindrically
symmetric solution at $r=r_0$. This means that we want the $(r,\eta)$-part of the metric to look like that of a
circle at $r=r_0$, i.e.
\begin{equation}
 \left.\frac{1}{f(r)}\left(dr^2 + b(r) f(r) d\eta^2\right)\right\vert_{r=r_0} \stackrel{!}{=} d\rho^2 + \rho^2 d\theta^2  \ ,
\end{equation}
where $(x,y)=\rho(\cos\theta,\sin\theta)$. We now replace $\rho^2=c_1(r-r_0)$ and $\theta=c_2\eta$, where $c_1$ and $c_2$ are 
constants that have to be determined. Using the expansion of the metric functions close to $r_0$ with
$f(r)=f'(r_0)(r-r_0)+...$ and $b(r)=b'(r_0)(r-r_0)+...$ we find that $c_1=4/f'(r_0)$ and $c_2=\sqrt{b'(r_0) f'(r_0)}/2$.
Now requiring that $c_2\eta=c_2\eta + 2\pi$ we find that the period of $\eta$ should be
\begin{equation}
 \tau_{\eta}=\frac{4\pi}{\sqrt{b'(r_0) f'(r_0)}} \ .
\end{equation}
Note that $T=\tau_{\eta}^{-1}(r_0\rightarrow r_h)$ is equal to the temperature of the corresponding black string solution (see (\ref{t_bs})).
Since the black string space-time possesses a compactified coordinate that has period $2\pi$, we will choose
$\tau_{\eta}=2\pi$ when constructing the phase diagrams.

\subsection{Boundary conditions}
The equations of motion have to be solved subject to appropriate boundary conditions. 
For the black string solution the boundary conditions for the metric functions at the horizon $r=r_h$ are 
\begin{equation}
 f(r_h)=0 \ \ , \ \ b(r_h)=0  \ 
\end{equation}
with $p(r_h)$ and $g(r_h)$ positive and finite. 
The requirement of regularity of the matter fields on the horizon leads to the following conditions
\begin{equation}
\label{bc_bs1}
\phi(r_h) =-g(r_h) A(r_h) \  \ ,
\end{equation}
\begin{eqnarray}
& &
-\alpha \left.\left[r^3 m^2\left(b'  \psi^2 +  (A')^2
  \psi  g^2 + 2   A'  \phi'  \psi  g +   \phi'^2  \psi \right)
+ r A^2 \left(2  b'  \psi^2  \psi' 
+   (A')^2  \psi  g^2
 + 2   A'  \phi'  \psi  g +    \phi'^2  \psi \right) \right]\right\vert_{r=r_h}\nonumber \\ 
&=&
\left.\left( r^5 m^2  (g')^2  \psi + 
 \frac{6 r^3}{l^2}  b'  \psi' - r^3  g'^2  A^2  \psi - 2  r^2 m^2  b'  \psi - 2   
A^2  b'  \psi  \right)\right\vert_{r=r_h}  \ , 
\end{eqnarray}
\begin{eqnarray}
& &\alpha \psi^2\left[r^2 m^2(r^2  g'  A'    g +  r^2  
g'  \phi'   - A' b') + r^2 A(
  2   g'  A  A'   g + \left. 2   g'  A  \phi'  -  
  A'^2   g^2 - 2  A'  \phi'    g  
-     \phi'^2)   - 2  A^2  A'  b'  \right] \right\vert_{r=r_h}  \nonumber \\
& =& \left(\left. r^4  g'^2  A  \psi^2 + 2 r  A  b'  \psi^2 + 
\frac{6 r^4}{l^2}  g'  A'  g + 6 r^4
 \frac{1}{l^2}  g'  \phi' - 6 r^2 \frac{1}{l^2}  A'  b' \right)\right\vert_{r=r_h} \ .
\end{eqnarray}
For the solitonic, cigar-shaped solutions the boundary conditions at $r=r_0$ are
\begin{equation}
f(r_0)=0 \ \ , \ \  b(r_0)=0  
\end{equation}
with $p(r_0)$ and $g(r_0)$ positive and finite.
For the matter functions the regularity conditions read
\begin{equation}
A(r_0)=\phi(r_0) g(r_0) \ \ , 
\end{equation}
\begin{eqnarray}
& &\left. \left(2b' \phi \psi^2-\frac{3 r}{\ell^2}  g' A' p+\frac{3 r}{\ell^2} g' p \phi' g-\frac{3 r}{\ell^2} b' \phi'\right)\right\vert_{r=r_0} \nonumber 
\\ &=& \left. \alpha\psi^2 \left[m^2 \left(
 - r g' A' p + r g' p \phi' g- r  b' \phi'\right) + r A'^2 \phi -2 r 
      A' \phi \phi'  g+ r \phi \phi'^2  g^2\right] \right\vert_{r=r_0}  \ ,
\end{eqnarray}
\begin{eqnarray}
& & \left.\left(-\frac{6 r}{\ell^2} p b' \psi'+2  m^2 p b'\psi-2  b' \phi^2 \psi \right)\right\vert_{r=r_0} \nonumber \\
&=&
\left. \alpha \left[m^2 r p \left(
 -2 b' \psi^2 \psi'+  A'^2  \psi
      -2   A'  \phi' \psi g+  \phi'^ 2 \psi g^2\right) + \phi^2 r\left(
      - A'^2 \psi + 2 A'  \phi' \psi g-  
       \phi'^2 \psi g^2\right)\right]\right\vert_{r=r_0}  \ .
\end{eqnarray}
\\

For both the black string and the solitonic solution the matter fields on the AdS boundary have the following behaviour
\begin{eqnarray}
\label{infty}
      \psi(r \gg 1) = \frac{\psi_-}{r} + \frac{\psi_+}{r^2} \ \ , \ \ 
      \phi(r \gg 1) = \mu - \frac{Q_e}{r} \ \ , \ \  
      A(r \gg 1) = \sigma - \frac{Q_m}{r} \ .
\end{eqnarray}
$\mu$ denotes the chemical potential, while $\sigma$ corresponds to the superfluid velocity. $Q_e$ and $Q_m$ denote the
electric and magnetic charge, respectively. In the following, we study the solutions with $\psi_-=0$. The dual operator ${\cal O}$
that corresponds to $\psi_+$ has mass dimension $\Delta=2$. 

The metric functions have the following asymptotic behaviour
\begin{eqnarray}
\label{infty2}
 f(r\gg 1)&=&\frac{r^2}{\ell^2} + \frac{\left(c_t + c_z\right)}{r} + O(r^{-2}) \ \ , \ \ 
b(r\gg 1)= \frac{r^2}{\ell^2} + \frac{c_t}{r} + O(r^{-2}) \ \ , \nonumber \\
p(r\gg 1)&=&\frac{r^2}{\ell^2} + \frac{c_z}{r} + O(r^{-2}) \ \ , \ \   g(r\gg 1) \sim O(r^{-3}) \ ,
\end{eqnarray}
where $c_t$ and $c_z$ are constants. Suitable combinations of these constants give the energy and tension per unit length
of the solutions \cite{mrs}.

\section{Solutions without scalar hair}
There are analytic solutions to the equations of motion in the case $\psi(r)\equiv 0$.  Here and in the following
we will set $\ell\equiv 1$. 

\subsection{Black string solutions}
In order to present the known black string
solutions, we use the radial coordinate $x$ that is related to $r$ by the following transformation
\begin{equation}
 r^2 = x^2 + a^2\frac{x_h^3}{x}  \ .
\end{equation}
Note that this transformation is perfectly regular for $ x > 0$. $a$ is a constant that is interpreted as a rotation parameter (see below)
and $x_h$ corresponds to the location of the (outer) horizon in the new coordinates with 
\begin{equation}
 r_h= \xi x_h  \ \ , \ \ \ \xi=\sqrt{1+a^2} \ .
\end{equation}

Uncharged and charged black string
solutions without scalar hair have been studied in \cite{lemos1,lemos2} and \cite{lemos3,dehghani3,dehghani2}, respectively, using the
coordinates $(x,\varphi,z)$.  
The solutions read
\begin{equation}
\label{bs1}
\phi(x)=\mu - \frac{Q}{x} \ \ \ , \ \ A(x)=\sigma + \frac{a Q}{\xi x}  \ ,
\end{equation}
\begin{equation}
\label{bs2}
ds^2 = -N(x) \left(dt+ L(x) d\varphi\right)^2 + K(x) d\varphi^2 + \frac{1}{F(x)} dx^2 + x^2 dz^2  \ , 
\end{equation} 
where 
\begin{eqnarray}
\label{bs3}
     N(x)= x^2-\frac{M\xi^2}{x}+\frac{\alpha \xi^2 Q^2}{2 x^2} \ \ & , & \ \ 
F(x)  =  x^2-\frac{M}{x}+\frac{\alpha Q^2}{2 x^2}    \ \ ,  \ \ L(x) = \frac{\xi a}{N(x) x}\left
(M-\frac{\alpha Q^2}{2x}\right)  \ \ , \nonumber \\
K(x) &=& x^2 + \frac{a^2 M}{x} - \frac{a^2 \alpha Q^2}{2 x^2} + N(x)L^2(x)  \ .
\end{eqnarray}
$Q$ denotes the charge per unit length of the solution.
These black string solutions have an outer horizon at $x=x_h$ and an inner horizon at $x=x_{-}< x_h$ with $F(x_{h})=0$, $F(x_-)=0$ provided that
$M \ge M_{\rm crit}=2^{5/4} \left(\alpha/3\right)^{3/4} Q^{3/2}$ \cite{dehghani3}. For $M=M_{\rm crit}$ the solution is an extreme
black string solution. If the black string is uncharged there is only one horizon at $x_h$.
For $a=0$ we have non--rotating black strings with $N=F$, $L=0$, $K = x^2$ and $A(x)\equiv \sigma$. Note that
for $a=0$ the coordinate $r$ is equal to the coordinate $x$. In this limit these solutions
are equivalent to the planar black hole solutions frequently used in the study
of holographic superconductors and superfluids in 2+1 dimensions. The only difference
is that for the planar black holes the boundary theory lives on $\mathbb{R}^3$, while
here it lives on $\mathbb{R}^2\times S^1$. Hence one dimension is compactified.

The mass ${\cal M}$ and angular momentum ${\cal J}$ per unit length are \cite{dehghani3}
\begin{equation}
 {\cal M}=\frac{1}{8}\left(3\xi^2 -1\right)M \ \ , \ \ {\cal J}=\frac{3}{8}\xi M a  \ .
\end{equation}

The temperature $T$ and entropy $S$ of the solution is given by \cite{dehghani3}
\begin{equation}
 T=\frac{3x_h}{4\pi \xi } - \frac{\alpha Q^2}{8\pi \xi x_h^3} 
=\frac{3r_h}{4\pi \xi^2 } - \frac{\alpha Q^2\xi^2}{8\pi r_h^3}=\frac{3x_h}{4\pi \xi } - 
\frac{\alpha \xi^3}{8\pi x_h}\left(\mu+\frac{a}{\xi}\sigma\right)^2   
\end{equation}
and
\begin{equation} 
S=\frac{\pi \xi x_h^2}{2}= \frac{\pi r_h^2}{2\xi}  \ .
\end{equation}
 
\subsection{Solitonic solutions}
Next to solutions which possess horizons, there are also solutions to the model that correspond
to solitonic solutions. For $g(r)\equiv 0$, i.e. in the static case and without gauge fields $\phi(r)\equiv 0$, $A(r)\equiv 0$ the 
solution corresponds to the AdS soliton discussed in \cite{witten2,horowitz_myers}. The metric reads
\begin{equation}
\label{analytic_cigar}
      f(r) = r^2 - \frac{r_0^3}{r} \ \ , \ \ p(r) = r^2 \ \ , \ \ b(r) = f(r)  \ \ , \ \ g(r)\equiv 0
\end{equation} 
obeying the  conditions $b(r_0)=0$, $f(r_0)=0$ for some radius $r_0$.
The form of the space-time doesn't change if we add a chemical potential $\phi(r)=\mu$. It was shown in \cite{nrt} that
at zero temperature and
for $\mu$ large the AdS soliton becomes unstable to the formation of scalar
hair at some critical value of the chemical potential $\mu_{cr}$. In the dual description this is interpreted as an insulator/superfluid phase transitions.
For small $\mu$, on the other hand, there is a phase transition from the AdS soliton to the corresponding black string 
for sufficiently large temperature \cite{ssw}. This corresponds to an insulator/fluid phase transition.
Note that in addition we can add a superfluid velocity to the model with $A(r)\equiv \sigma$. This will not change the
space-time.

In contrast to the black string case no solution to the full coupled set of equations of motion seems
to exist for $\psi\equiv 0$. However, we managed to find a solution for the electric potential in the background of the
solution (\ref{analytic_cigar}), i.e. for  
$\alpha=0$ and letting $g(r)\equiv 0$, $A(r)\equiv 0$. This reads
\begin{equation}
      \phi(r) = C_0 + C_1 \left[ 
                                \log\left( \frac{\left(r-1\right)^2}{r^2+r+1}\right) + 2 \sqrt{3} {\rm atan}\left(\frac{2r+1}{\sqrt 3} \right)
      \right] \ ,
\end{equation}
where $C_0$ and $C_1$ are integration constants. Asymptotically this solution behaves like 
\begin{equation}
 \phi(r\gg 1) = C_0 + \sqrt{3} \pi C_1 - \frac{6C_1}{r} + O(r^{-2}) \ .
\end{equation}
Hence, we can interpret $C_0+\sqrt{3} \pi C_1$ as chemical potential $\mu$ and $6C_1$ as electric
charge $Q_e$ in the context of this paper. Note that the equivalent solution for a scalar potential in the
background of a black string metric is given by $\phi(r)=\mu - Q_e/r$ (see also (\ref{bs1})). Hence in the background
of a black string the electric field possesses a simple monopole charge, while in the background of the
AdS soliton it carries additional multipoles.


\section{Free energy}

The free energy is given by $\Omega = T \tilde{S}$ \cite{hks}, where
$\tilde{S}$ is the euclidean version of the action (\ref{action}) and the tilde now and in the following denotes
the evaluation on-shell.
However, this integral will in general be divergent and suitable counterterms have to be added \cite{balasubramanian}. 
These counterterms depend on the curvature invariants
of the induced metric on the boundary. 
The total action that determines the free energy then reads \cite{hhh}
\begin{equation}
S_{\rm total}=S_E + S_{\rm ct} + S_{\psi}
\end{equation}
where
$S_E$ is the action given in (\ref{action}) and 
\begin{equation}
 S_{\rm ct}= \left. \int d^3 x \sqrt{-g_{\infty}} \left(-2K + 4\right) \right\vert_{r=r_{\infty}}
\end{equation}
is the counterterm action containing a Gibbons-Hawking term and a boundary cosmological constant.
$K=g^{\mu\nu} \nabla_{\mu} n_{\nu}$ denotes the extrinsic
curvature and $g_{\infty}$ is the metric evaluated on the AdS boundary $r=r_{\infty}$.
Moreover, for a scalar field that behaves asymptotically like $\psi\rightarrow \psi_{+}/r^2$ we have to add the following
term \cite{hhh}
\begin{equation}
 S_{\psi}= -\int d^3 x \sqrt{-g_{\infty}} \left.\left( 2\psi n^{\mu} \partial_{\mu} \psi + \psi^2
\right) \right\vert_{r=r_{\infty}}  \ .
\end{equation}
Now, our metric is of the form
\begin{equation}
 ds^2 = g_{tt}(r) dt^2 + g_{rr}(r) dr^2 + 2g_{t\chi}(r) dtd\chi + g_{\chi\chi}(r) d\chi^2 + g_{zz}(r) dz^2 \ 
\end{equation}
with $\sqrt{-g}=\sqrt{-(g_{tt}g_{\chi\chi}-g_{t\chi}^2) g_{rr} g_{zz}}$.
$\chi$ is a periodic coordinate with $\chi=\varphi$ for black strings and $\chi=\eta$ for cigar solutions.

To compute $\tilde{S}_{E}$ we 
use the equations of motion and the fact that $T_z^z= {\cal L}_m$. 
We can then express the action density in terms of $R_z^z$ by 
\be
    R + \frac{6}{\ell^2} + 2\alpha {\cal L}_{m} = 2 R_z^z   \ .
\ee
Moreover, we observe that $R_z^z$ can be expressed as a total derivative as follows
\begin{equation}
 \sqrt{-g} R_z^z = \frac{d}{dr}\left(\frac{1}{2}\sqrt{-g} g^{zz} g^{rr} \frac{dg_{zz}}{dr}\right)  \ .
\end{equation}
Since $g^{rr}$ vanishes at the horizon $r=r_h$ and the cigar tip $r=r_0$, respectively, the value
of $T \tilde{S}_E /V_2$, where $V_2$ corresponds to the 2-dimensional volume associated to the coordinates
$(\chi, z)$, is equal to  the value of the expression in the brackets evaluated on the AdS boundary.


To compute the counterterm, we note that
the outward pointing unit vector $n^{\mu}$ on the AdS boundary is here given by
$n_{\mu} dx^{\mu} = \sqrt{g_{rr}} dr$.
We then obtain
\begin{eqnarray}
\frac{T(\tilde{S}_{\rm ct} + \tilde{S}_{\psi})}{V_2} = \left. \sqrt{g_{tt} g_{zz} g_{\varphi\varphi}}
\left( -\psi^2 + 4 -2 \frac{1}{\sqrt{(g_{tt}g_{\chi\chi}-g_{t\chi}^2) g_{rr} g_{zz}}}
\partial_r \left(\sqrt{(g_{tt}g_{\chi\chi}-g_{t\chi}^2) g_{zz}}\right) - \frac{2}{\sqrt{g_{rr}}}\psi \partial_r 
\psi\right) \right\vert_{r=r_{\infty}}  \ . 
\end{eqnarray}

Using these results and the expansion of the matter and metric functions at infinity (\ref{infty}), (\ref{infty2}) 
we can express $\Omega$  in terms of the coefficients $c_t$, $c_z$ as follows
\be
\label{fe1}
          \left(\frac{\Omega}{V_2}\right)_{\rm (H)BS} = c_t - 2 c_z
\ee
for black strings solutions with (HBS) and without (BS) scalar hair and 
\be
\label{fe2}
          \left(\frac{\Omega}{V_2}\right)_{\rm (H)C} = c_t +  c_z  
\ee
for solitonic, cigar-shaped solutions with (HC) and without (C) scalar hair, respectively.

\section{Numerical results}
In the following we will choose $m^2 = -2$ for all our calculations which is well above the Breitenlohner-Freedman bound of
$m^2_{\rm BF}=-9/4$ \cite{bf}. For all our numerical computations we have used the ODE solver COLSYS \cite{colsys}.
The relative errors of the solutions are on the order of $10^{-10}$ - $10^{-6}$. 

\subsection{Fluid/Superfluid phase transitions}
The gravity dual to a holographic fluid/superfluid phase transitions is the formation of scalar hair
on a black hole solution. In our case, we will study scalar hair formation on the horizon of a black string.

For $\alpha=0$ these solutions have been studied in detail in \cite{hartmann_brihaye}. The scalar-gauge field equations
were solved in the background of the uncharged ($Q=0$) black string solution (\ref{bs2}), (\ref{bs3}).
It has been found that the order of the phase transition changes from being 2nd order to being 1st order if
the ratio between the superfluid velocity and chemical potential $\sigma/\mu$ and/or the rotation parameter
$a$ are sufficiently large. Here, we want to investigate what happens away from the probe limit, i.e.
we will take backreaction of the space-time into account. 

It has been noticed previously that the inclusion of backreaction decreases the critical
temperature at which condensation sets in. We also see this here.
Our results for different choices of $g(r_h)$ and $\alpha$ are given in Fig.\ref{fig1}, where
we give the dependence of the critical temperature $T_c$ on the ratio between superfluid velocity $\sigma$ and chemical
potential $\mu$. For a given $\sigma/\mu$ the critical temperature $T_c$ is decreasing with increasing
$\alpha$ which is clearly seen when comparing the curves for $\alpha=0.01$ (black) and those for $\alpha=0.2$ (red).
While in \cite{hartmann_brihaye} and other studies on holographic superfluids, the ratio $\sigma/\mu$ was always chosen to be
positive, we have extended our results to negative values of $\sigma/\mu$ here. We observe that the curves
are mirror-symmetric with respect to a value of $\sigma/\mu=(\sigma/\mu)_{\rm s}$ that is negative and for a fixed value of $g(r_h)$ seems
to be nearly independent of $\alpha$. Moreover, the larger the value of $g(r_h)$ the smaller $(\sigma/\mu)_{\rm s}$.
We also observe that for $g(r_h)=0$  the value of $(\sigma/\mu)_{\rm s}=0$, i.e. the equations are
invariant under the exchange $\sigma/\mu \rightarrow -\sigma/\mu$. This is related to the fact that
the equations are in general invariant under the exchange $(g(r),A(r))\rightarrow (-g(r),-A(r))$. But
the equations are symmetric under $A(r)\rightarrow -A(r)$ only for the choice $g(r_h)=0$ (see also the boundary condition (\ref{bc_bs1})).

 \begin{figure}[htp]
\hbox to\linewidth{\hss%
	\resizebox{9cm}{7cm}{\includegraphics{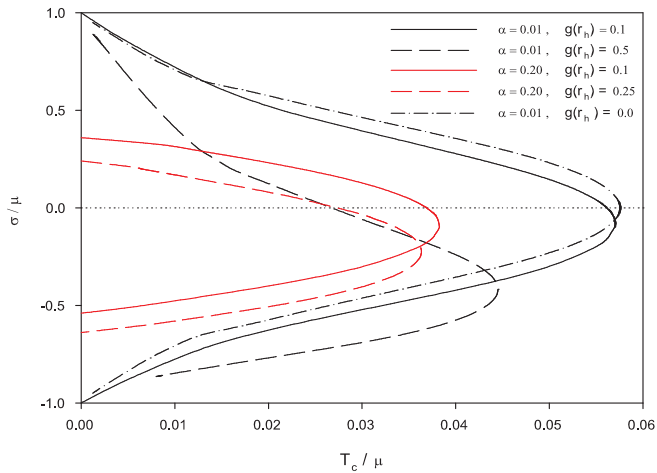}}
\hss}
\caption{{\small The value of the critical temperature $T_c/\mu$ is shown in dependence
on the ratio between superfluid velocity $\sigma$ and chemical potential $\mu$ for different
choices of $\alpha$ and $g(r_h)$. 
}
 }
\label{fig1}
\end{figure}

Without backreaction it was found that for sufficiently large angular momentum and/or ratio
of the superfluid velocity $\sigma$ and the chemical potential $\mu$ the phase transition is always
1st order. Here we find that if the backreaction is strong enough the phase transition
becomes again 2nd order. This is shown in Fig.\ref{fig2} for $\sigma/\mu=0.2$, $g(r_h)=0.1$ and
two different values of $\alpha$. For $\alpha=0.1$ the phase transition is 1st order,
while it is 2nd order for $\alpha=0.3$. This is also confirmed when considering the
free energy, see Fig.\ref{fe_sigma04_a01}. For $\alpha=0.1$, the free energy is non-continuous at
the phase transition, while it is continuous for $\alpha=0.3$. Moreover, for $\alpha=0.3$ the free energy of the
superfluid phase is always lower than that of the fluid phase, while for $\alpha=0.1$ there exists a temperature
interval in which the free energy of the superfluid phase is larger than that of the fluid phase.

 \begin{figure}[ht]
\hbox to\linewidth{\hss%
	\resizebox{10cm}{6cm}{\includegraphics{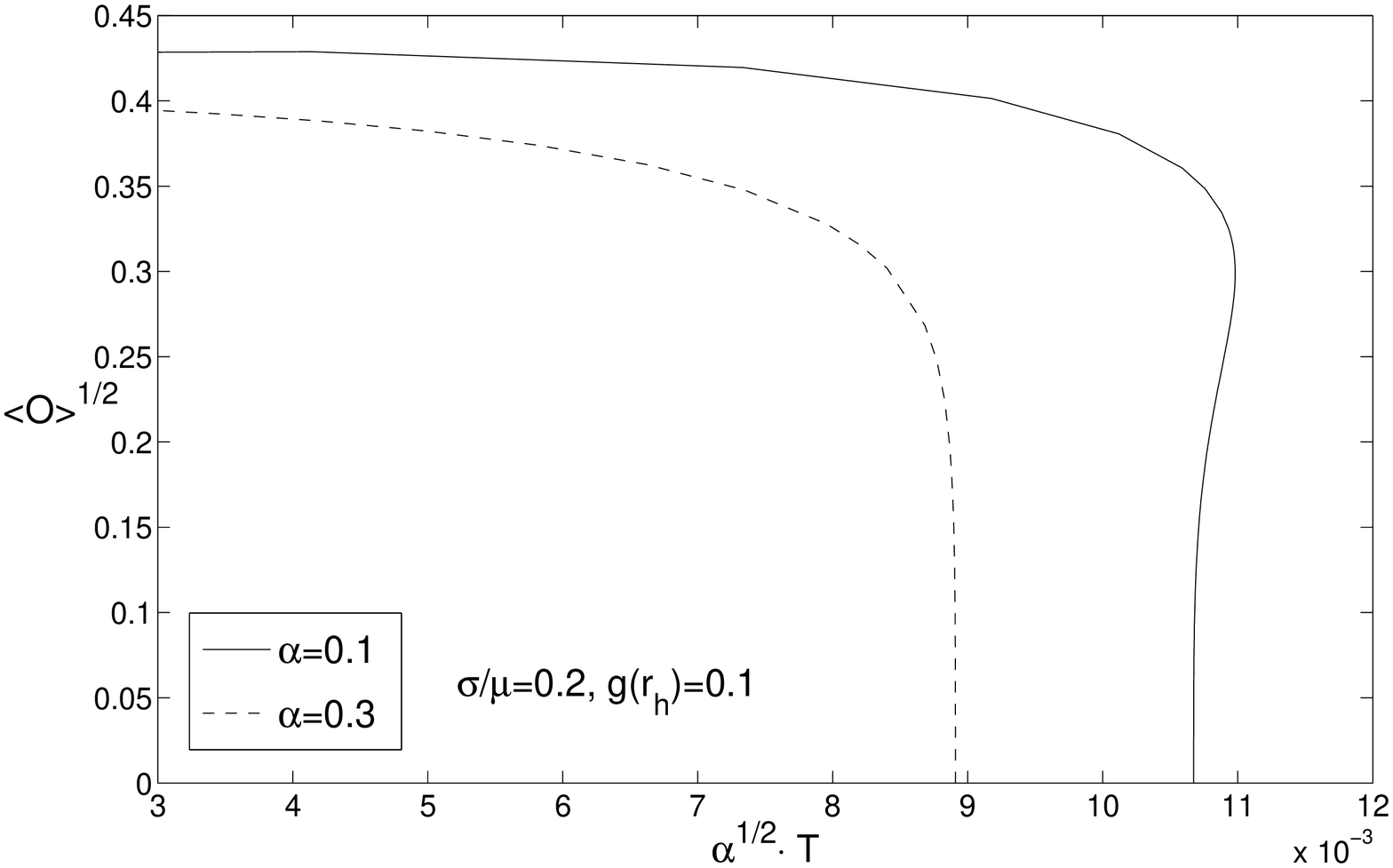}}
\hss}
\caption{{\small The condensate $<{\cal O}>^{1/2}\equiv\psi_+^{1/2}$ as function of the temperature $T$  
for $\sigma/\mu=0.2$, $g(r_h)=0.1$ and two different values of the gravitational coupling $\alpha$.
}
 }
\label{fig2}
\end{figure}

 \begin{figure}[ht]
\hbox to\linewidth{\hss%
	\resizebox{10cm}{6cm}{\includegraphics{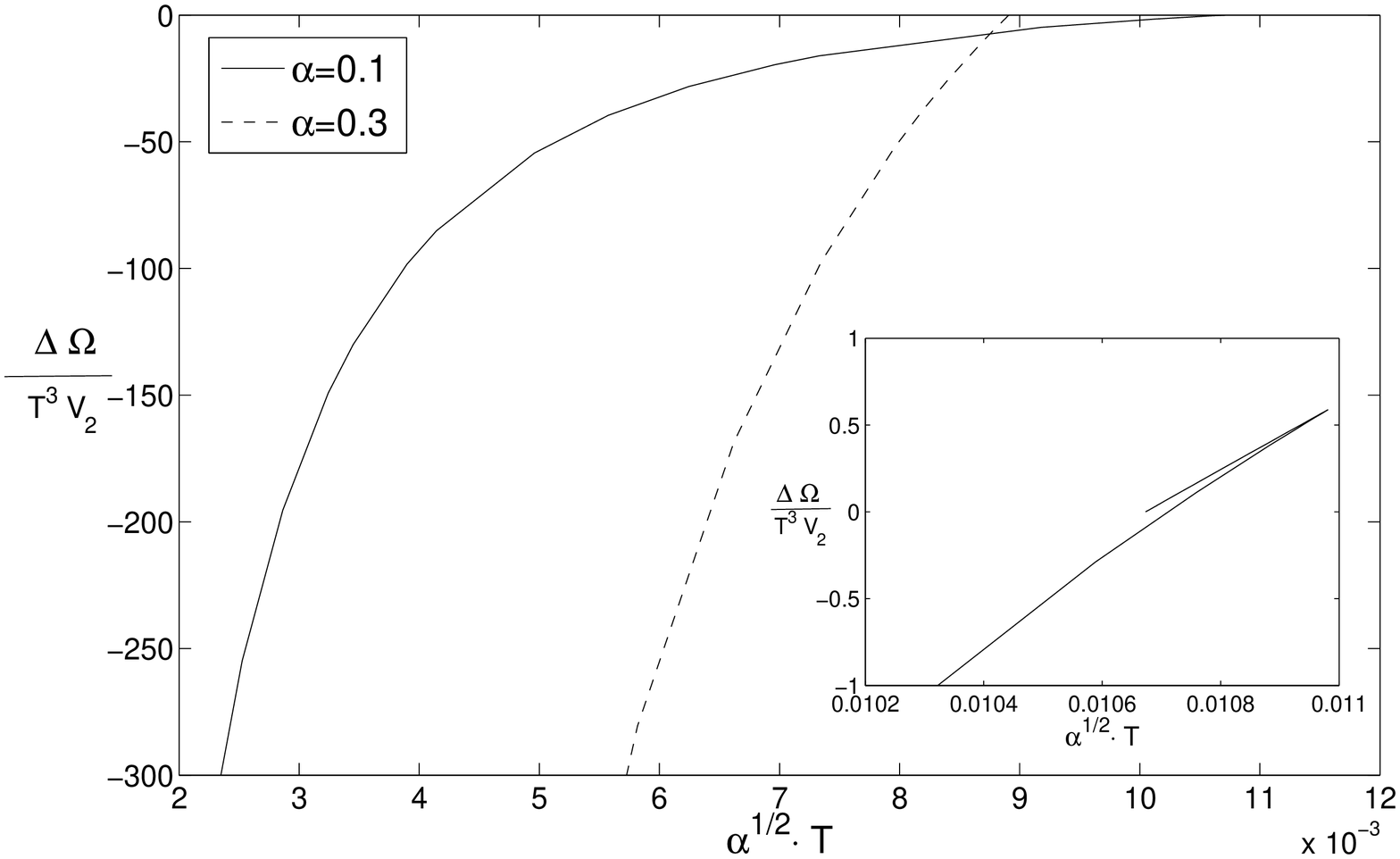}}
\hss}
\caption{{\small The difference $\Delta\Omega/(T^3 V_2)$ of the free energy of the superfluid phase and of the
fluid phase  as function of the temperature $T$  
for $\sigma/\mu=0.2$, $g(r_h)=0.1$ and two different values of the gravitational coupling $\alpha$.
}
 }
\label{fe_sigma04_a01}
\end{figure}

We have then investigated how the order of the phase transition depends on the ratio between superfluid velocity $\sigma$ and chemical
potential $\mu$ as well as on the horizon velocity $g(r_h)$. Our results are given in Fig.\ref{order_bs}.
We give the critical value of $\sigma/\mu$ at which the phase transition changes its order in dependence
on $g(r_h)$. We observe that for $\sigma=0$ the phase transition is always second order. Increasing $\sigma/\mu$ to sufficiently large values
or decreasing $\sigma/\mu$ to sufficiently small values the phase transition becomes 1st order. This was also
observed in the probe limit \cite{hartmann_brihaye}. Increasing $\alpha$ enlarges the domain in the $\frac{\sigma}{\mu}$-$g(r_h)$-plane in which
the phase transition is 2nd order. 

\begin{figure}[ht]
\hbox to\linewidth{\hss%
	\resizebox{10cm}{6cm}{\includegraphics{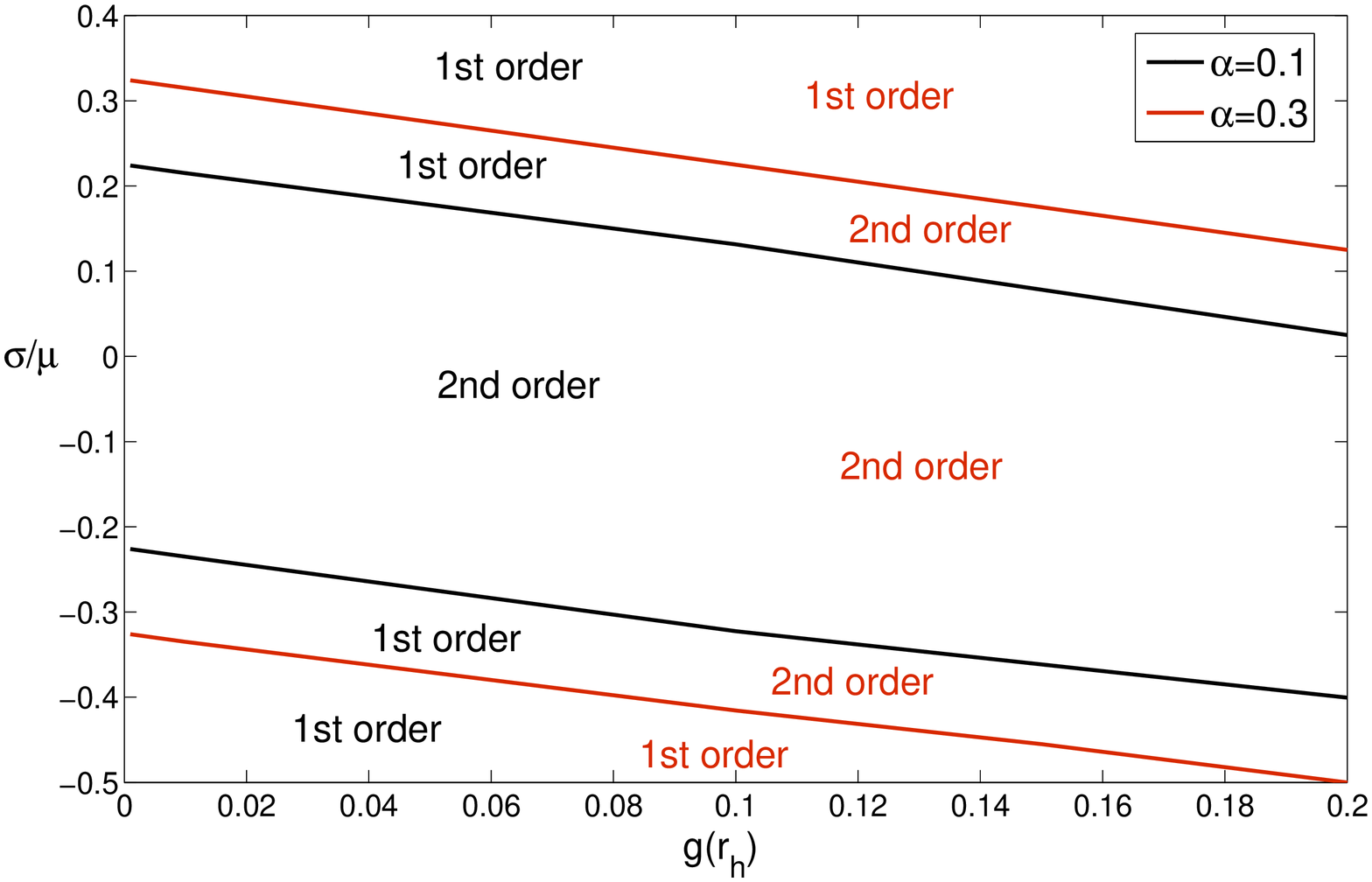}}
\hss}
\caption{{\small The value of $\sigma/\mu$ at which the phase transition changes its order is given
as function of $g(r_h)$ for $\alpha=0.1$ (black) and $\alpha=0.3$ (red). 
}
 }
\label{order_bs}
\end{figure}

\subsection{Insulator/Superfluid phase transitions}
The gravity dual of an insulator/superfluid phase transition is the formation of scalar hair on an AdS soliton, in our
case at the tip of a solitonic, cigar-shaped solution. 
Like black string solutions, cigar solutions are characterized by the chemical potential $\mu$
and the value of the condensate $\psi_+$. 
In the case $\alpha=0$, $g(r)\equiv 0$, $A(r)\equiv 0$ the equations for the matter fields read
\be
\label{cigar_alpha_0}
     (f \phi')' = \phi\psi^2 \ \ , \ \ (r^2 f \psi')' = m^2 r^2 \psi - \phi^2 \psi
\ee
with $f$ given in (\ref{analytic_cigar}).
The asymptotic form of the solution to (\ref{cigar_alpha_0}) (with $r_0=1$) then is
\be
     \phi(r\gg 1) = \mu - \frac{Q_e}{r} + \frac{\mu \psi_+^2 - 3 Q_e}{12} \frac{1}{r^4} + O(r^{-5}) \ \ , \ \ 
     \psi(r\gg 1) = \psi_+ \left(\frac{1}{r^2} - \frac{ \mu^2}{6 r^4}\right) + O(r^{-4})  \ . 
\ee

In the case $\alpha > 0$ the matter equations are still of the form (\ref{cigar_alpha_0})
but the Einstein equations for the metric functions cannot be solved explicitly. Hence, the solutions
are only available numerically.
Let us first discuss solutions with $g(r_0)=0$ and vanishing superfluid velocity, $\sigma=0$. Note that
this choice will lead to $g(r)\equiv 0$ and the space-time is static. This case corresponds to the
(3+1)-dimensional analogue of the solutions studied in \cite{horowitz_way}.   
We find that for a fixed value of the backreaction parameter $\alpha$ the soliton solution starts to 
form scalar hair for $\mu > \mu_{cr}$. For $\mu < \mu_{cr}$ we have $\psi(r)\equiv 0$, $\phi(r)\equiv \mu$
and the metric functions given by (\ref{analytic_cigar}). This is shown in Fig. \ref{cigar2} where
we give the value of the condensate $\psi_+^{1/2}=<{\cal O}>^{1/2}$, the value of the metric
function $p(r)$ at the tip $r=r_0$ of the soliton and the difference between the value of the
scalar potential $\phi(r)$ at infinity, which is equal to the chemical potential $\mu$, and the value of 
$\phi(r)$ at $r=r_0$. In agreement with the results in \cite{horowitz_way} we find that $\mu_{cr}$ does not depend on the strength of backreaction,
i.e. on $\alpha$. For static solutions, we find that $\mu_{cr} \approx 0.859$ for all values of $\alpha$. 
Moreover, we observe that at the critical value $\mu_{cr}$ the difference $\mu-\phi(r_0)$ tends to zero indicating
that $\phi(r)\equiv \mu$ in this limit. We also find that when taking backreaction into
account that $p(r_0)$ becomes very small for large values of $\mu$. The solution seems to tend to a black string solution, where
the period $\tau_{\eta}$ decreases strongly approaching this limit. Since we are only interested in the solutions
with $\tau_{\eta}=2\pi$ for the phase diagrams we didn't consider this limit in more detail and leave this as future work.

Now let us turn to the case where the soliton solution rotates around the $z$-axis.
In this case $g(r)$ is non-vanishing and $A(r)$ is non-trivial and possesses a
non-vanishing value $\sigma$ on the AdS boundary, which we interpret as
superfluid velocity. 
Our results for this case are shown in Fig.\ref{cigar3}, where
we give the value of the condensate $\psi_+^{1/2}=<{\cal O}>^{1/2}$, the value of the metric
function $p(r)$ at the tip $r=r_0$ of the soliton and the difference between the value of the
scalar potential $\phi(r)$ at infinity, which is equal to the chemical potential $\mu$, and the value of 
$\phi(r)$ at $r=r_0$ for $\alpha=1$, $g(r_0)=0.5$ and different values of $\sigma$. We observe that the critical 
value $\mu_{cr}$ depends on $\sigma$ and in fact increases with increasing $\sigma$.
The limiting solution does not have a constant scalar potential $\phi(r)$, but possesses
a non-vanishing, positive charge $Q_e$.

For the construction of the phase diagrams we will later choose $\tau_{\eta}=2\pi$. However, in order to understand how the choice
of period influences the properties of the solutions we have investigated the dependence of
$r_0$, the critical value of the chemical potential $\mu_{cr}$  at which condensation sets in
as well as of the electric charge $Q_e$ of the limiting solution at $\mu=\mu_{cr}$ 
on the value of the period for two different values of $g(r_0)$. This is shown in Fig.\ref{fig_period_cigar}.
Note that $Q_e=0$ for $g(r_0)=0$ since for this choice the limiting solution corresponds to
the AdS soliton (\ref{analytic_cigar}) with constant $\phi(r)=\mu$. This is different for $g(r_0)=0.5$. In this
case, the limiting solution is not given by a constant  scalar potential, but by a potential
of the form $\phi(r\gg 1)=\mu - \frac{Q_e}{r}$, where $Q_e$ is negative. We observe that all quantities
decrease with increasing period and with increasing $g(r_0)$.

\begin{figure}
\centering
\epsfysize=8cm
\mbox{\epsffile{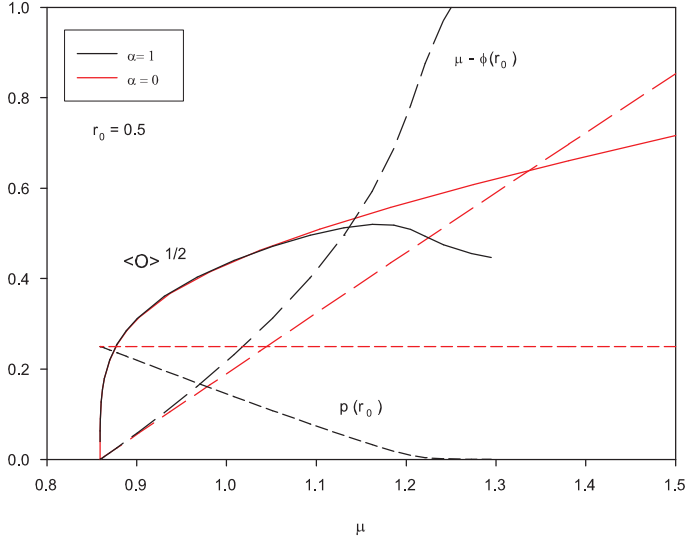}}
\caption{\label{cigar2}
The condensate $<{\cal O}>^{1/2}=\psi_+^{1/2}$, the value of the metric function $p(r)$ at 
$r=r_0$ as well as the difference of the chemical potential $\mu=\phi(\infty)$ and the value of the scalar
potential $\phi(r)$ at $r=r_0$ is shown in dependence on $\mu$ 
for solitonic, cigar-shaped solutions with $r_0=0.5$  and for two values of $\alpha$. Here $g(r_0)=0$ and $\sigma=0$. }
\end{figure}

\begin{figure}
\centering
\epsfysize=8cm
\mbox{\epsffile{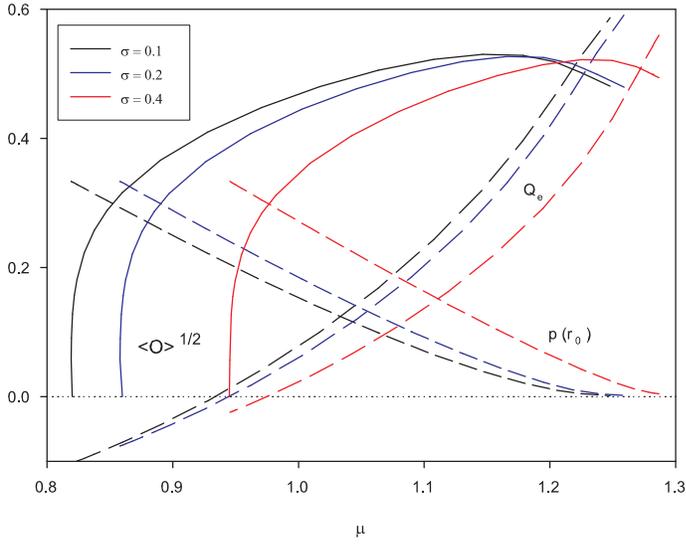}}
\caption{\label{cigar3}
The condensate $<{\cal O}>^{1/2}=\psi_+^{1/2}$, the value of the metric function $p(r)$ at 
$r=r_0$ as well as the difference of the chemical potential $\mu=\phi(\infty)$ and the value of the scalar
potential $\phi(r)$ at $r=r_0$ is shown in dependence on $\mu$ 
for $\alpha=1$, $g(r_0)=0.5$ and different values of $\sigma$. }
\end{figure}

\begin{figure}
\centering
\epsfysize=8cm
\mbox{\epsffile{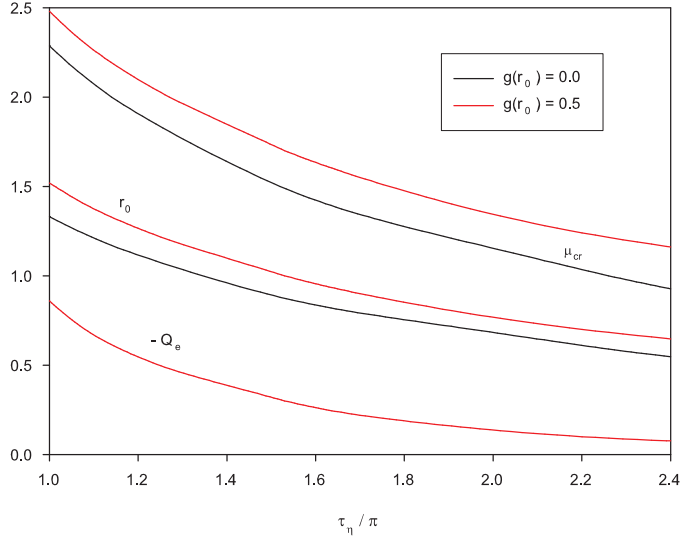}}
\caption{\label{fig_period_cigar}
The critical chemical potential $\mu_{cr}$ at which condensation sets in, 
the radius $r_0$ at the tip of the cigar as well as the electric charge $Q_e$ of the solution at $\mu=\mu_{cr}$ is shown in dependence
on the period $\tau_{\eta}$ for $g(r_0)=0$, $\sigma=0$ and for $g(r_0)=0.5$, $\sigma=0.4$, respectively.
Here $\alpha=0.2$. Note that $Q_e\equiv 0$ for $g(r_0)=0$.}
\end{figure}


\subsection{Phase diagrams}
We can now put all the information together and construct phase diagrams that will tell us
which phase has the lowest free energy and is hence the preferred phase for a given value of
$T$, $\mu$ and $\sigma$. We will first discuss the case of non-rotating gravity solutions with vanishing
superfluid velocity $\sigma=0$, which correspond to the (3+1)-dimensional analogues of the solutions
studied in \cite{horowitz_way}. We will then discuss the influence of the superfluid velocity on the
phase diagram. 

In order to be able to compare the free energy of the black string and of the cigar solutions,
we fix the period $\tau_{\eta}=2\pi$. This in turn will fix the value of $r_0$. For the non-rotating
solitonic, cigar-shaped solutions without hair, i.e. with $\psi\equiv 0$, we have that 
\be
\label{period_cigar}
            \frac{\tau_{\eta}}{\pi} = \frac{4}{\sqrt{f'(r_0) b'(r_0)}}  = \frac{4}{3 r_0}  \ .
\ee
Hence, we choose $r_0=2/3$ which gives 
\begin{equation}
\left(\frac{\Omega}{V_2}\right)_{\rm C}=-r_0^3=(-2/3)^3 \approx -0.29 
\end{equation}
for the non-rotating solitonic solutions without scalar hair and period $\tau_{\eta}=2\pi$. 
Note that this value is independent of the value of the chemical potential $\mu$ and of the value of the superfluid velocity $\sigma$
since for non-rotating solitonic solutions without scalar hair we have $\phi(r)\equiv \mu$ and $A(r)\equiv \sigma$. 

For non-rotating black strings without scalar hair we have
\begin{equation}
 T=\frac{3 r_h}{4\pi}\left(1- \frac{\alpha \mu^2}{6 r_h^2}\right) \ \ , \ \ \left(\frac{\Omega}{V_2}\right)_{\rm BS}= - r_h^3 
\left(1 + \frac{\alpha \mu^2}{2r_h^2}\right)  \ .
\end{equation}
Here, the superfluid velocity $\sigma$ does not enter since for non-rotating black strings without scalar hair we have $A(r)\equiv \sigma$.
Using the expression for the temperature $T$ we can rewrite $\Omega$ in terms of $\mu$ and $T$. Then comparing $\Omega_{\rm BS}$
and $\Omega_{\rm C}$ we find the phase boundary between cigar (C) and black string (BS) solutions, i.e.
between holographic insulators and fluids. Note that for $T=0$ this is given by $\mu=\frac{2}{3}4^{-1/3}\sqrt{6}\alpha^{-1/2}\approx 1.0287 \alpha^{-1/2}$.
For $\mu=0$ this transition is at $T=1/(2\pi)$. This is natural since here the periodicity of Euclidean time $1/T$ becomes
equal to the period $\tau_{\eta}$ of the $\eta$-coordinate. 

For solitonic and black string solutions which are only known numerically we have used the formulae (\ref{fe1}) and (\ref{fe2}), respectively.
Note that when taking a non-vanishing superfluid velocity into account the solutions without scalar hair are still
static, i.e $g(r)\equiv 0$ with the choice $g(r_h)=0$ and $g(r_0)=0$, respectively. This changes, however, when
considering solutions with scalar hair. As soon as $\psi(r)\neq 0$ the choice $\sigma\neq 0$ together with $g(r_h)=0$ (resp. $g(r_0)$)
does {\it not} lead to $g(r)\equiv 0$. In other word: while the solutions with scalar hair,  non-vanishing superfluid
velocity and $g(r_h)=0$ (resp. $g(r_0)=0$) have neither a rotation on the AdS boundary nor on the horizon $r_h$ (resp. tip $r_0$)
$g(r)$ is non-vanishing for $r\in ]r_h : \infty [$ (resp.  $r\in ]r_0 : \infty [$).

The free energy of the different non-rotating solutions  
as function of the temperature $T$ is given in Fig. \ref{freeenergy} for $\alpha=0.2$ and two different value $\mu$.
At the critical temperature $T_c$ at which $\psi_+=0$ the free energy of the black string fluid (BS) and that of the
black string superfluid (HBS) become equal as expected. For small $T$ the free energy of the soliton superfluid (HC) is the lowest
and hence dominates. Increasing the temperature, the free energy of the black string fluid decreases until at some
sufficiently high $T=T_d$ the free energy of the black string fluid becomes lower than that of the soliton superfluid.
In the dual theory this means that for $T < T_d$ the superfluid dominates, while for $T > T_d$ it is the fluid.
Note that for $\mu=2$ we find that $T_c < T_d$. This is different for $\mu=2.5$. In this case, the value of $T_c$ is slightly above 
$T_d$ such that there is a small temperature interval $ T_d < T < T_c$ in which the free energy of the
hairy black string is the lowest. In the dual theory this means that for $ T < T_d$ the  soliton superfluid is preferred,
while for  $ T_d < T < T_c$ it is the black string superfluid, while for $ T > T_c$ it is the
fluid.

\begin{figure}[p!]
  \begin{center}
    \subfigure[$\mu=0.2$]{\label{mu2}\includegraphics[scale=0.55]{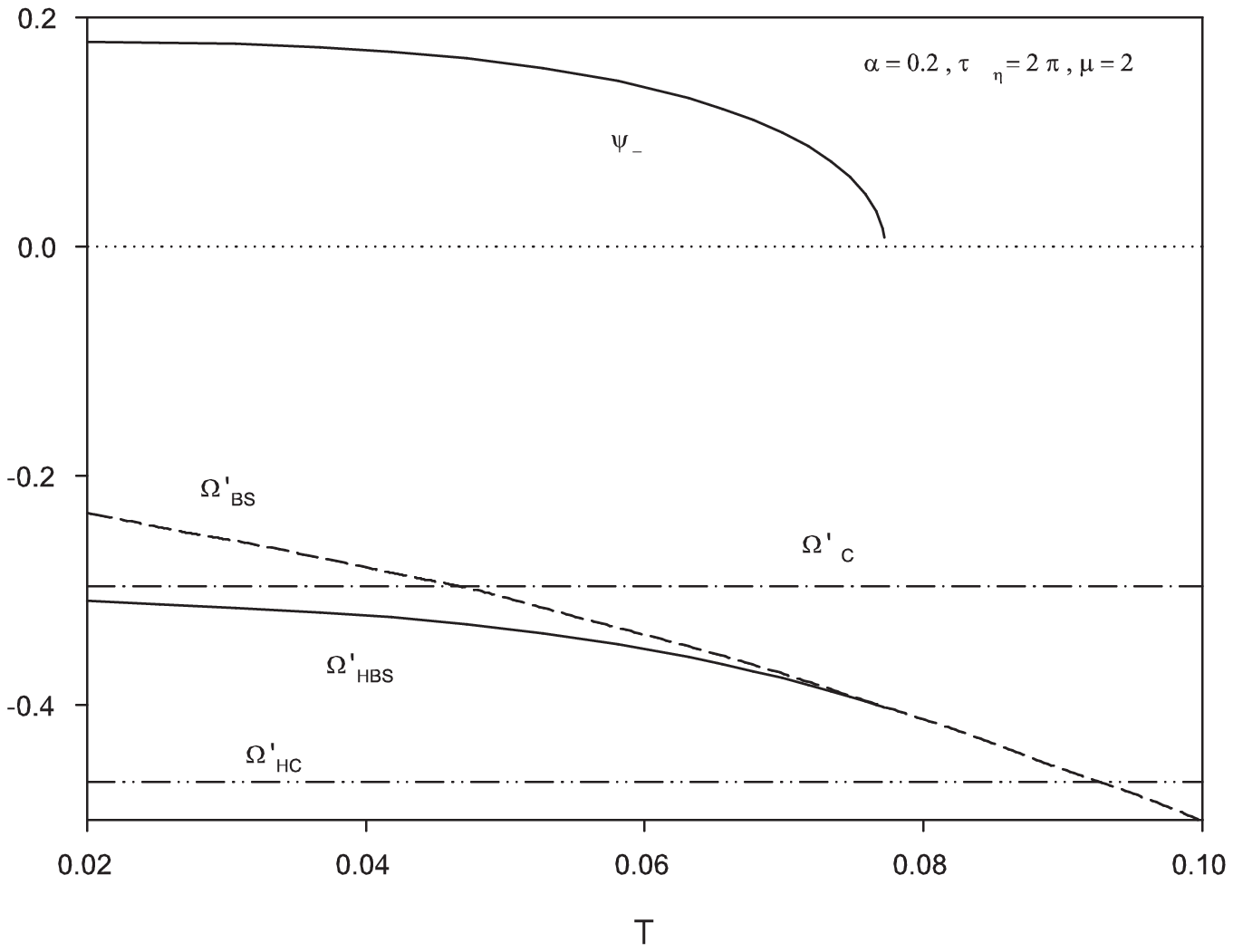}}
    \subfigure[$\mu=0.25$]{\label{mu25}\includegraphics[scale=0.55]{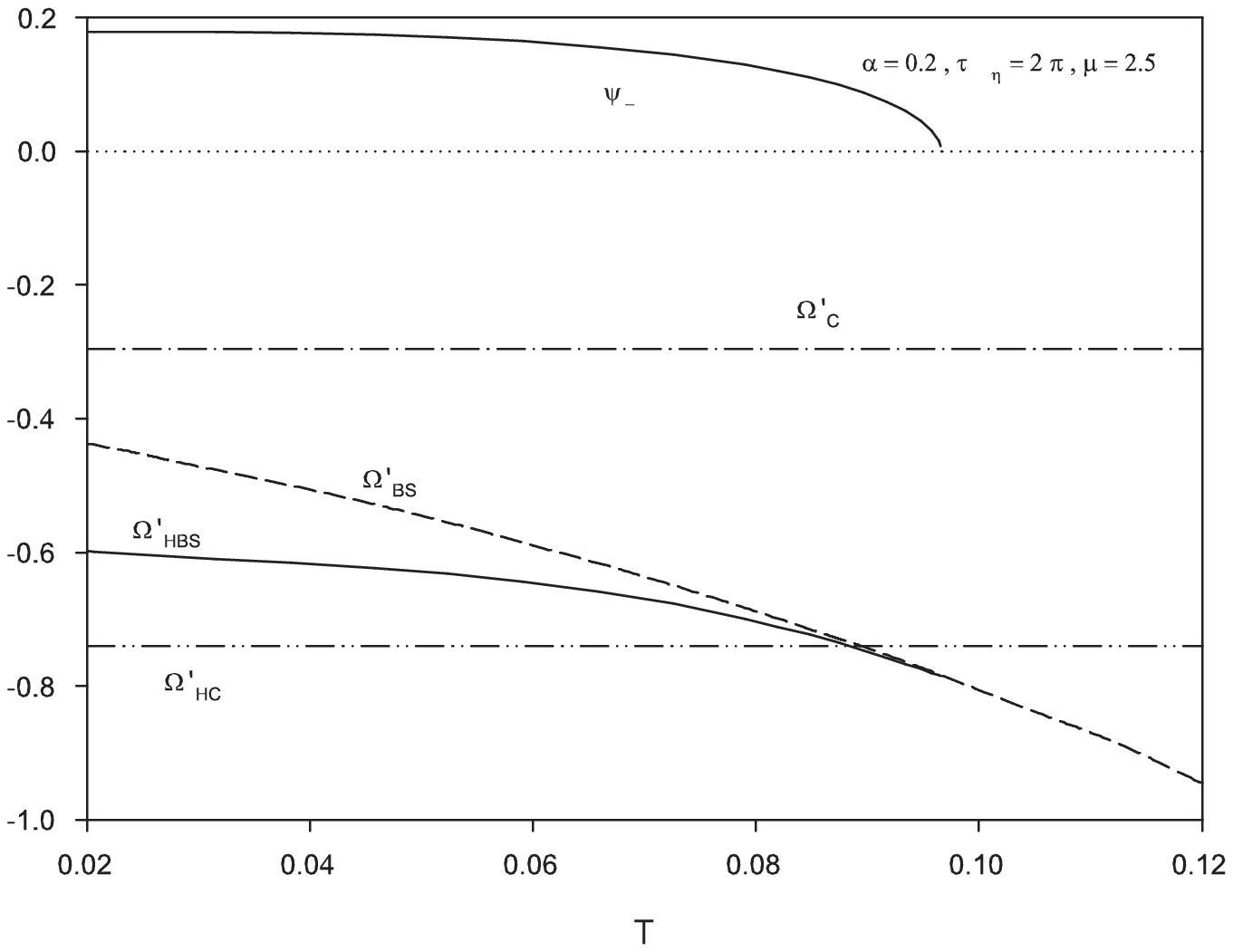}}
   
   \end{center}
   \caption{The condensate $\psi_+=<{\cal O}>$ developing on the black string is shown in dependence on $T$ for
$\alpha=0.2$, $\tau_{\eta}=2\pi$ and two different values of $\mu$.
Also shown is the free energy $\Omega'=\Omega/V_2$ of the black string solution (index BS), of the hairy black string solutions (index HBS),
of the cigar-shaped solution (index C) and of the hairy cigar-shaped solution (index HC), respectively. 
Note that the solution with the lowest free energy dominates.}  \label{freeenergy}
  \end{figure}

Our results for different values of $\alpha$, different $\sigma$ and $g(r_h)=g(r_0)=0$ are shown in Fig. \ref{phases}. 
We find that for small $\alpha$ there
are four phases present in the phase space diagram (see Fig.\ref{alp001}, Fig.\ref{alp02} and Fig.\ref{alp02_r}). In these cases
the preferred phase for small $T$ and small $\mu$ is the soliton, cigar-shaped phase (C) corresponding to an insulator.
For small $\mu$ and increasing $T$ there is a first order phase transition from a soliton to a black string (BS) \cite{ssw} 
corresponding to a soliton insulator/black string fluid phase transition. For small temperature and increasing chemical potential $\mu$ the cigar-shaped solution
becomes unstable to form scalar hair. This is a holographic soliton insulator/soliton superfluid phase transition. Finally for large
temperature and increasing $\mu$ the black string becomes unstable to the formation of scalar hair. This corresponds
to a black string fluid/black string superfluid phase transition. Note that for small $\alpha$ there are two triple points at which
three phases coexist. At $(\mu^*_{(1)}, T^*_{(1)})$ the soliton insulator, the soliton superfluid and the fluid coexist, while
at $(\mu^*_{(2)}, T^*_{(2)})$ the soliton superfluid, the fluid and the black string superfluid coexist. 
We observe that these two triple points move closer together with increasing $\alpha$, where
$T^*_{(1)}$, $\mu^*_{(2)}$ and $T^*_{(2)}$ decrease with $\alpha$, while $\mu^*_{(1)}=\mu_{cr}$ is independent of $\alpha$ (see discussion above).
For sufficiently large $\alpha$ the hairy cigar solution never dominates and has disappeared from the phase diagram (see Fig.\ref{alp08}).
If we choose $\mu$ slightly above  $\mu_{cr}=\mu^*_{(1)}$ and decrease the temperature we observe that the system goes
from a black string fluid to a black string superfluid and then to a soliton insulator. This is a new type of phase transition that has already been
observed in \cite{horowitz_way} in (4+1)-dimensional AdS. We find that for 
static solutions with $\sigma=0$ and $\alpha=0.8$ the range of chemical potential in which this
new phase transition exists is $1.1 \lesssim \mu \lesssim 1.15$. 

\begin{figure}[p!]
  \begin{center}
    \subfigure[$\alpha=0.01$, $\sigma=0$]{\label{alp001}\includegraphics[scale=0.55]{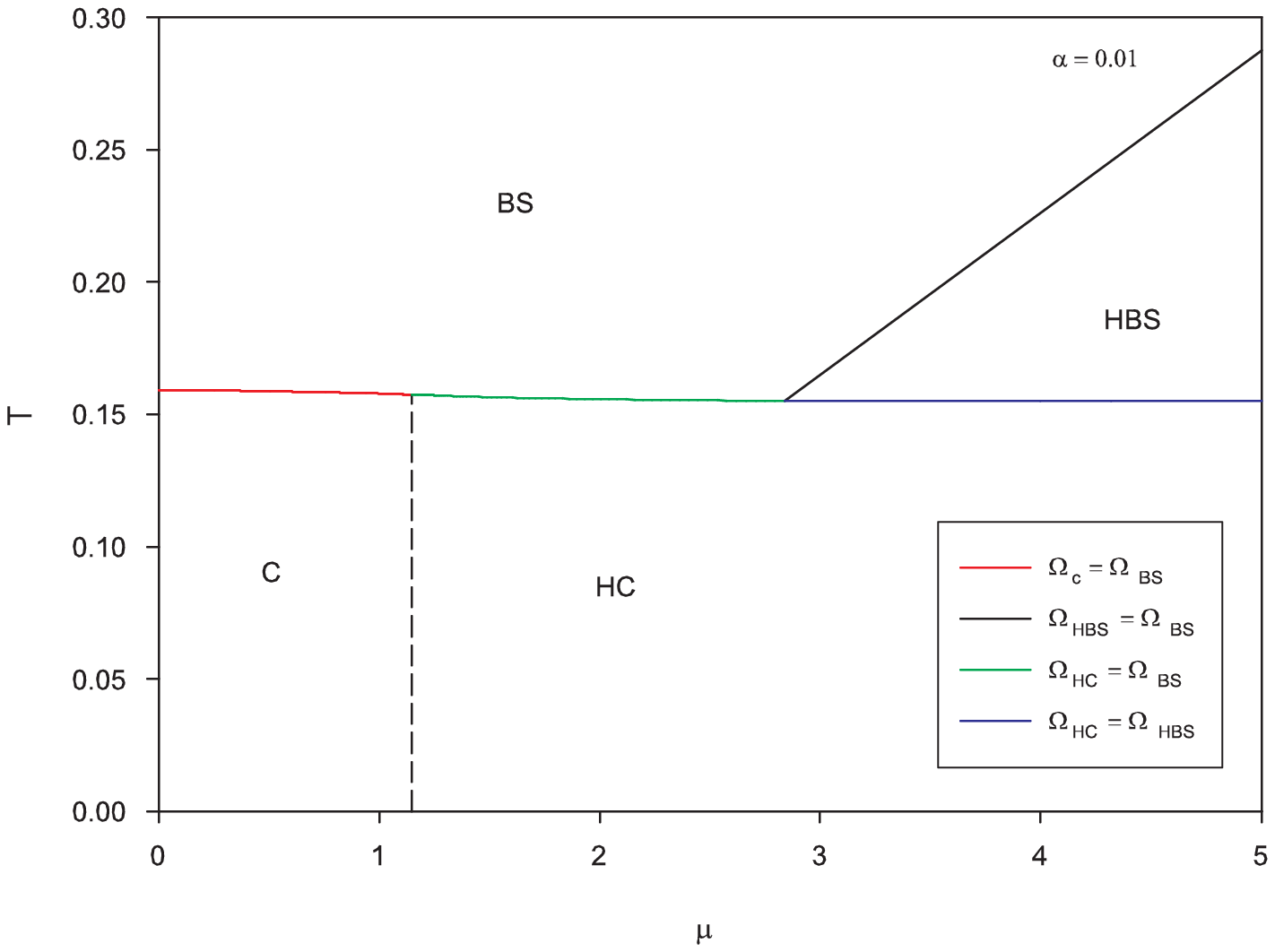}}
    \subfigure[$\alpha=0.2$, $\sigma=0$]{\label{alp02}\includegraphics[scale=0.55]{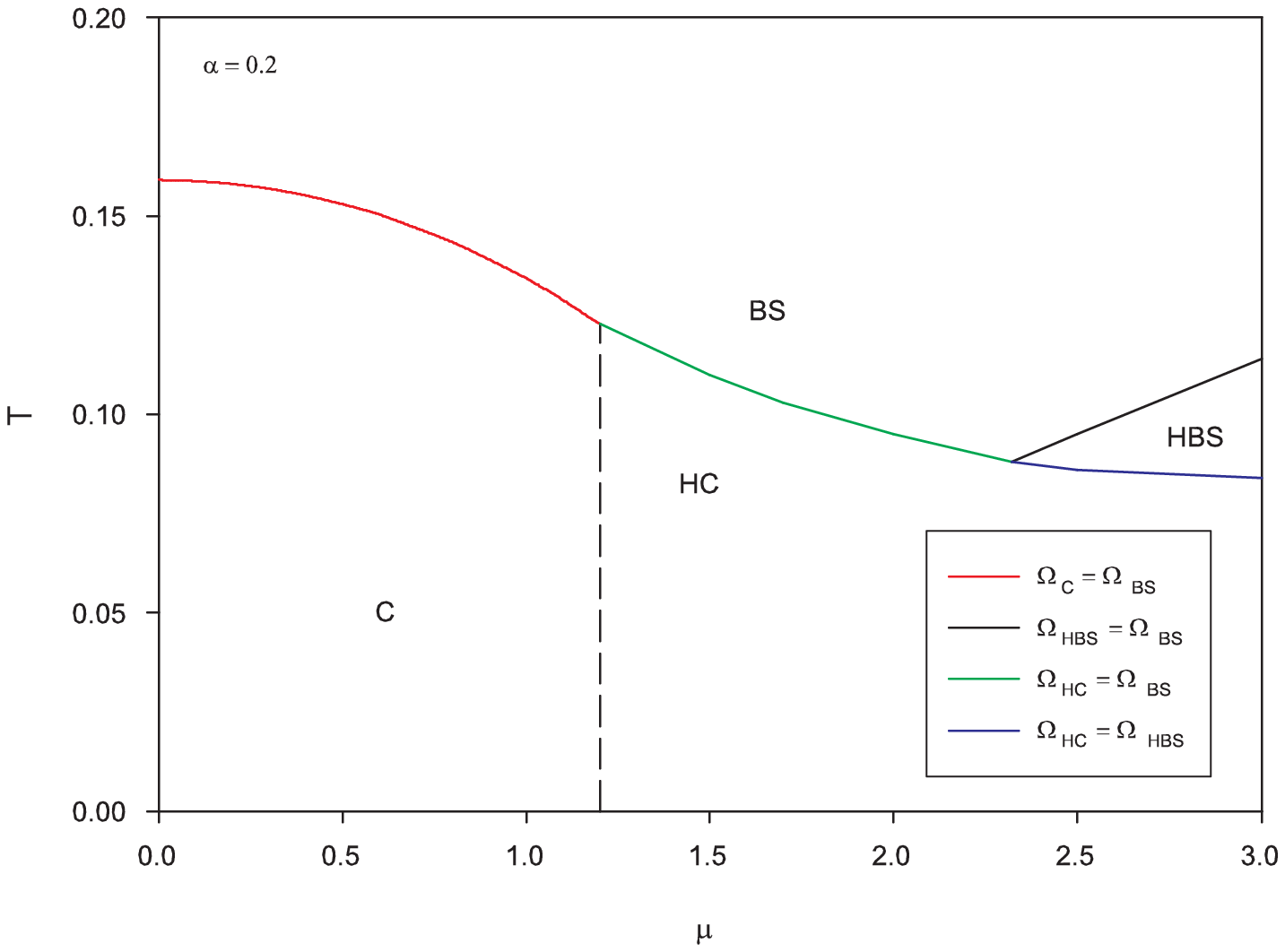}}\\
    \subfigure[$\alpha=0.8$, $\sigma=0$]{\label{alp08}\includegraphics[scale=0.55]{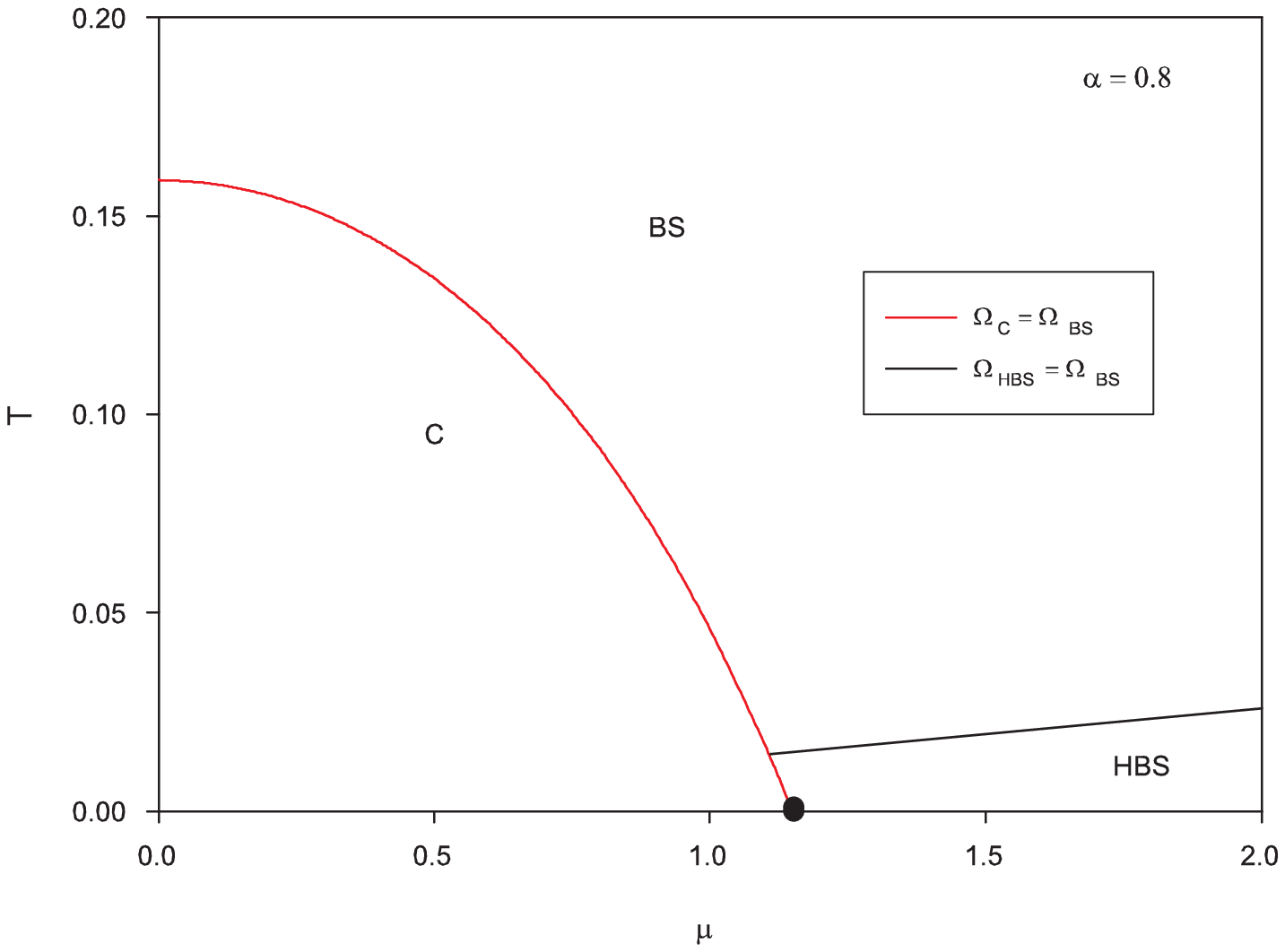}}
    \subfigure[$\alpha=0.2$, $\sigma=\mu/4$]{\label{alp02_r}\includegraphics[scale=0.55]{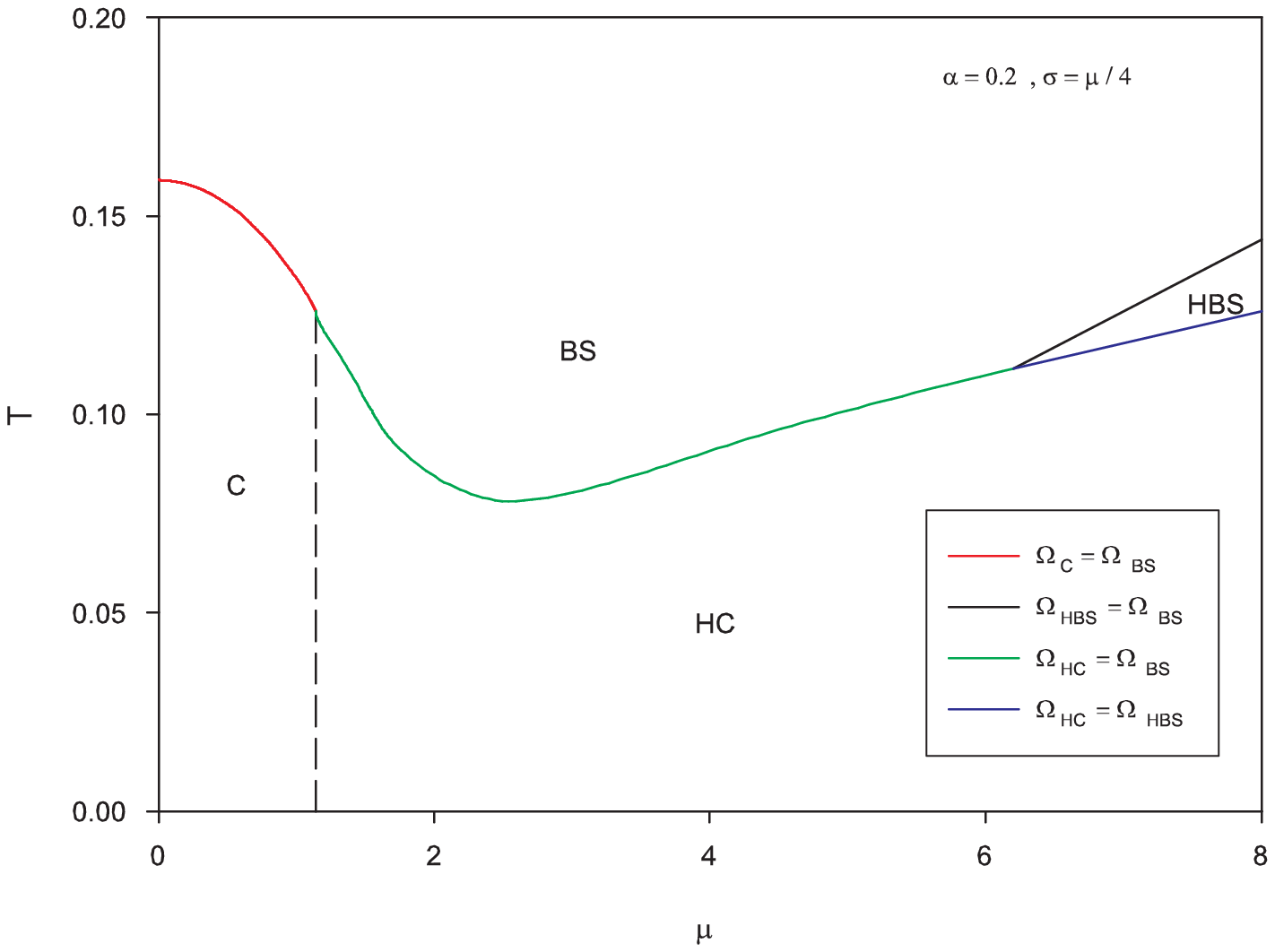}}  
   \end{center}
   \caption{Phase diagrams for $\tau_{\eta}/\pi = 2$ and different choices of the backreaction parameter $\alpha$ and the 
ratio between superfluid velocity
$\sigma$ and chemical potential $\mu$. The letters indicate the phase that is dominating with the abbreviations BS: black string, 
C: solitonic, cigar-shaped solution, HBS: black string with scalar hair and HC: solitonic, cigar-shaped solution
with scalar hair, respectively. The red, black, green and blue lines represent the values of
$T$ and $\mu$ at which $\Omega_{\rm C}=\Omega_{\rm BS}$, $\Omega_{\rm HBS}=\Omega_{\rm BS}$, $\Omega_{\rm HC}=\Omega_{\rm BS}$ and 
$\Omega_{\rm HC}=\Omega_{\rm HBS}$, respectively.
The dotted vertical line represents the value of $\mu_{cr}$, i.e. the value of
the chemical potential at which the solitonic solution develops scalar hair. This is temperature-independent.
In Fig. \ref{alp08} $\mu_{cr}$ is represented by the dot on the $T=0$ axis. For all solutions $g(r_h)=g(r_0)=0$. }  \label{phases}
  \end{figure}

The superfluid velocity has an important influence on the phase diagram as seen in Fig.\ref{alp02_r}, which
is for $\alpha=0.2$, $\sigma=\mu/4$ and $g(r_h)=g(r_0)=0$. We find that $\mu_{cr}$ and with it
the values $(\mu^*_{(1)}, T^*_{(1)})$ at the first triple point are independent
of $\sigma$ (within numerical accuracy). Moreover, the second triple point is at larger values of $\mu$ and 
$T$, i.e. $(\mu^*_{(2)},T^*_{(2)})$  increases with $\sigma$. In addition, for intermediate $\mu$ the fluid phase
dominates for smaller values of $T$ when increasing $\sigma$. In other words: while an increase of the superfluid velocity
does have practically no influence on the soliton insulator/black string fluid and the soliton insulator/soliton superfluid
phase transitions, it influences the soliton superfluid/black string superfluid and the black string fluid/black string superfluid
phase transitions considerably. In addition, we observe that a new type of phase transition is possible for non-vanishing
superfluid velocity: for $\alpha=0.8$ and fixed $0.08 \lesssim T \lesssim 0.11$  we can now have a transition from
a soliton insulator to a soliton superfluid to a black string fluid and then back to a soliton superfluid when increasing the chemical
potential. For vanishing
superfluid velocity the transition from a fluid to a soliton superfluid is not possible for fixed $T$ and increasing
$\mu$, while here it is. 

\section{Conclusions}

In this paper we have studied the gravity duals of superfluid/fluid/insulator phase transitions in (2+1)-dimensions.
We find that there are up to four phases corresponding in the gravity theory to black strings with and without scalar hair
and solitonic, cigar-shaped solutions with and without scalar hair, respectively. Taking backreaction into account we find that
the chemical potential at which the soliton insulator/soliton superfluid phase transition takes place
is independent of the strength of the backreaction and increases with increasing superfluid velocity.
While the phase transition from a black string fluid to a black string superfluid is 1st order for sufficiently
large superfluid velocity and small backreaction, it is 2nd order for large enough backreaction.
We observe that for small backreaction there are four phases. For small $\mu$ and small (resp. large) $T$ the 
soliton insulator (resp. black string fluid) dominates, while for larger $\mu$ and small (resp. large) $T$ the
soliton superfluid (resp. black string superfluid) dominates. For large enough backreaction the 
soliton superfluid never dominates and a new type of phase transition from a soliton insulator to a black string superfluid
exists in a small range of $\mu$ when increasing the temperature $T$. For non-vanishing
superfluid velocity we find a new type of phase transition from a black string fluid to a soliton superfluid
in a small range of $T$ and for increasing $\mu$.

In \cite{horowitz_way} the limit of vanishing charge of the scalar field which in our paper corresponds to $\alpha=\infty$ 
has also been studied. In this limit the scalar field cannot be made real by a U(1) gauge transformation.
The theory is effectively a gravity-complex scalar field model that has solutions in the form of planar boson
stars \cite{bs}. These have been interpreted to be the gravity dual of a Bose-Einstein condensate.
In \cite{horowitz_way} only the simplest, non-rotating boson stars have been considered. It would be interesting to
see how rotating and/or excited planar boson star configurations can 
be interpreted in the dual theory. \\
\\
{\bf Acknowledgments} YB thanks the Belgian FNRS for financial support and would like 
to acknowledge R. Manvelyan for discussions.

\begin{small}

\end{small}

\end{document}